\documentclass[journal=ancac3,manuscript=article]{achemso}
\usepackage{bm}
\usepackage{graphicx}
\usepackage{amssymb}
\usepackage{amsthm}
\usepackage{amsmath}
\usepackage{achemso}

\usepackage{hyperref}
\hypersetup{colorlinks=true, citecolor=blue, urlcolor=blue, linkcolor=blue}

\usepackage[version=3]{mhchem} 
\usepackage[T1]{fontenc}       
\usepackage{indentfirst}

\author{Xue Chen}
\affiliation{State Key Laboratory of Superlattices and Microstructures, Institute of Semiconductors, Chinese Academy of Sciences, Beijing 100083, China}
\alsoaffiliation{Center of Materials Science and Optoelectronics Engineering \& CAS Center of Excellence in Topological Quantum Computation, University of Chinese Academy of Sciences, Beijing, 100049, China}

\author{Sven Reichardt}
\affiliation{Physics and Materials Science Research Unit, University of Luxembourg, Luxembourg 1511, Luxembourg}

\author{Miao-Ling Lin}
\affiliation{State Key Laboratory of Superlattices and Microstructures, Institute of Semiconductors, Chinese Academy of Sciences, Beijing 100083, China}

\author{Yu-Chen Leng}
\affiliation{State Key Laboratory of Superlattices and Microstructures, Institute of Semiconductors, Chinese Academy of Sciences, Beijing 100083, China}

\author{Yan Lu}
\affiliation{State Key Laboratory of Superlattices and Microstructures, Institute of Semiconductors, Chinese Academy of Sciences, Beijing 100083, China}

\author{Heng Wu}
\affiliation{State Key Laboratory of Superlattices and Microstructures, Institute of Semiconductors, Chinese Academy of Sciences, Beijing 100083, China}
\alsoaffiliation{Center of Materials Science and Optoelectronics Engineering \& CAS Center of Excellence in Topological Quantum Computation, University of Chinese Academy of Sciences, Beijing, 100049, China}

\author{Rui Mei}
\affiliation{State Key Laboratory of Superlattices and Microstructures, Institute of Semiconductors, Chinese Academy of Sciences, Beijing 100083, China}
\alsoaffiliation{Center of Materials Science and Optoelectronics Engineering \& CAS Center of Excellence in Topological Quantum Computation, University of Chinese Academy of Sciences, Beijing, 100049, China}

\author{Ludger Wirtz}
\email{ludger.wirtz@uni.lu}
\affiliation{Physics and Materials Science Research Unit, University of Luxembourg, Luxembourg 1511, Luxembourg}

\author{Xin Zhang}
\email{zhangxin@semi.ac.cn}
\affiliation{State Key Laboratory of Superlattices and Microstructures, Institute of Semiconductors, Chinese Academy of Sciences, Beijing 100083, China}
\alsoaffiliation{Center of Materials Science and Optoelectronics Engineering \& CAS Center of Excellence in Topological Quantum Computation, University of Chinese Academy of Sciences, Beijing, 100049, China}

\author{Andrea C. Ferrari}
\affiliation{Cambridge Graphene Centre, University of Cambridge, 9 JJ Thomson Avenue, Cambridge CB3 0FA, UK}

\author{Ping-Heng Tan}
\email{phtan@semi.ac.cn}
\affiliation{State Key Laboratory of Superlattices and Microstructures, Institute of Semiconductors, Chinese Academy of Sciences, Beijing 100083, China}
\alsoaffiliation{Center of Materials Science and Optoelectronics Engineering \& CAS Center of Excellence in Topological Quantum Computation, University of Chinese Academy of Sciences, Beijing, 100049, China}

\title[An \textsf{achemso} demo]
  {Control of Raman scattering quantum interference pathways in graphene}

\abbreviations{GIC, RRS, REP, FWHM, EPME}
\begin{document}


\begin{abstract}
Graphene is an ideal platform to study the coherence of quantum interference pathways by tuning doping or laser excitation energy. The latter produces a Raman excitation profile that provides direct insight into the lifetimes of intermediate electronic excitations and, therefore, on quantum interference, which has so far remained elusive. Here, we control the Raman scattering pathways by tuning the laser excitation energy in graphene doped up to 1.05eV, above what achievable with electrostatic doping. The Raman excitation profile of the G mode indicates its position and full width at half maximum are linearly dependent on doping. Doping-enhanced electron-electron interactions dominate the lifetime of Raman scattering pathways, and reduce Raman interference. This paves the way for engineering quantum pathways in doped graphene, nanotubes and topological insulators.
\end{abstract}
\section{Introduction}
Interference between quantum pathways can occur in all physical systems\cite{Quantum-Interference-and-Coherence}, as demonstrated by electron collisions\cite{1998-Nature-QI}, conductance jumps\cite{2019-NC-perovskite}, transmission dips\cite{Guedon-2012-NN,single-2019-NM,Greenwald-2021-NN}, exciton transports\cite{Glazov-PRL-2020}, magnetoconductance\cite{Stojetz-PRL-2005}, chemical reaction dynamics\cite{crQI-2020-science} and inelastic light scattering\cite{wangfeng2011Nature,SWNT-2012-PRL,2020-ACSNano-Fano}. In the quantum picture of the Raman scattering process\cite{Light-Scattering-in-Solids}, incident photons (energy $E_{\rm L}$) induce electronic excitations, which then generate phonons, followed by the radiation of scattered photons. The intermediate electronic excitations act as quantum pathways, thus they can interfere with each other. As $E_{\rm L}$ is tuned to approach the electronic transition of interest, resonant Raman scattering (RRS) occurs\cite{Light-Scattering-in-Solids}, greatly enhancing some quantum pathways\cite{Sven-2017-PRB}, therefore the Raman intensity\cite{Basko-NJP09-Theo,wangfeng2011Nature}. The ability to control quantum pathways provides a unique opportunity to detect\cite{wangfeng2011Nature}, understand\cite{Liu-2013-nanolett,SWNT-2012-PRL} and exploit\cite{NC-2022-ReS,2020-ACSNano-Fano} inelastic light scattering, and to design quantum interference-based devices\cite{Guedon-2012-NN,single-2019-NM,2019-NC-perovskite,Greenwald-2021-NN}.

The effect of quantum interference on the intensity of Raman modes was reported in silicon\cite{CdS-1970-PRL}, CdS\cite{CdS-1970-PRL}, carbon nanotubes\cite{SWNT-2012-PRL,DWNT-2012-NC}, graphene\cite{Dmode-2004-PPB,Basko-NJP09-Theo, enhance-2010-acsnano,wangfeng2011Nature,Ferrari-2013-NN,Liu-2013-nanolett,Hasdeo-PRB-2016,Sven-2017-PRB}, MoTe$_2$\cite{nanoph-2012-MoTe,Sven-2017-nanolett} and ReS$_2$\cite{NC-2022-ReS}. The unique band structure of single layer graphene (SLG) makes it ideal to study quantum interference\cite{Dmode-2004-PPB,Basko-NJP09-Theo,wangfeng2011Nature,Ferrari-2013-NN,Liu-2013-nanolett,Hasdeo-PRB-2016,Sven-2017-PRB}, as it enables continuous control of the Raman scattering pathways, by tuning the electrostatic doping, $E_{\rm F}$, relative to a fixed $E_{\rm L}$\cite{Basko-NJP09-Theo,enhance-2010-acsnano,wangfeng2011Nature,Ferrari-2013-NN,Liu-2013-nanolett}, or conversely, tuning $E_{\rm L}$ under a fixed $E_{\rm F}$\cite{FeCl3-2019-PRM}. Such $E_{\rm F}$ or $E_{\rm L}$ handle, paves the way to optical control of intermediate electronic excitations\cite{wangfeng2011Nature}. Raman experiments in SLG showed enhanced G and D peak intensities for $E_{\rm F}$ approaching $0.5E_{\rm L}$\cite{wangfeng2011Nature,Liu-2013-nanolett}. The enhancement was limited to 7 in Refs.\cite{enhance-2010-acsnano,Liu-2013-nanolett}, due to defects and inhomogeneous dopants introduced by ionic gel dielectrics. We previously showed that FeCl$_3$-intercalation into graphite can produce SLG flakes without defects with $E_{\rm F}$ up to $\sim$-1eV\cite{zhao2011FeCl3}, achieving a high carrier concentration $\sim7.56\times10^{13}$ cm$^{-2}$. Here, we use this to control the intermediate electronic excitations by tuning $E_{\rm L}$ for heavily-doped SLG with fixed $E_{\rm F}$.

Performing Raman measurements by tuning $E_{\rm L}$ produces the so-called Raman excitation profile (REP), i.e. a plot of intensity of Raman modes as a function of $E_{\rm L}$\cite{Light-Scattering-in-Solids}. This allows one to directly monitor the lifetimes of intermediate electronic excitations, because the REP width is a signature of electronic energy broadening, i.e. it is proportional to the inverse lifetime of Raman scattering pathways\cite{Eb-2011-PRBPedro}. These can be modified by electron-electron (e-e) interactions\cite{ee-2007-PRB,Basko-2009-PRB} and electron-phonon (e-ph) coupling\cite{epc-2008-PRB}. Their effect on Raman scattering pathways and overall quantum interference have not been investigated thus far, to the best of our knowledge.

Here, we control quantum interference of Raman scattering pathways in SLG by tuning $E_{\rm L}$ in hole (h)-doped SLG produced by intercalating FeCl$_3$ into graphite\cite{zhao2011FeCl3}. The G peak REP features a single resonant peak, whose position depends linearly on $E_{\rm F}$. The full width at half maximum, FWHM, of the G peak REP also depends linearly on $E_{\rm F}$. We show that such doping-induced REP broadening is determined by the Raman scattering pathways lifetime, and is dominated by e-e interactions, enhanced by carrier concentration. This shows that $E_{\rm L}$ can be used to control the allowed Raman scattering pathways in SLG. In principle, quantum interference is always present in Raman scattering from any materials, with the exact form of e-e and e-ph interactions depending on their electronic structure. Thus, besides SLG and topological insulators with Dirac-like band structure, one can expect to control Raman scattering pathways in intercalated or substitution-doped few-layer graphene, metallic nanotubes, and anisotropic layered materials.

\section{Results and discussion}
\subsection{Raman spectroscopy of doped graphene}
\begin{figure*}[tb]
\centerline{\includegraphics[width=160mm]{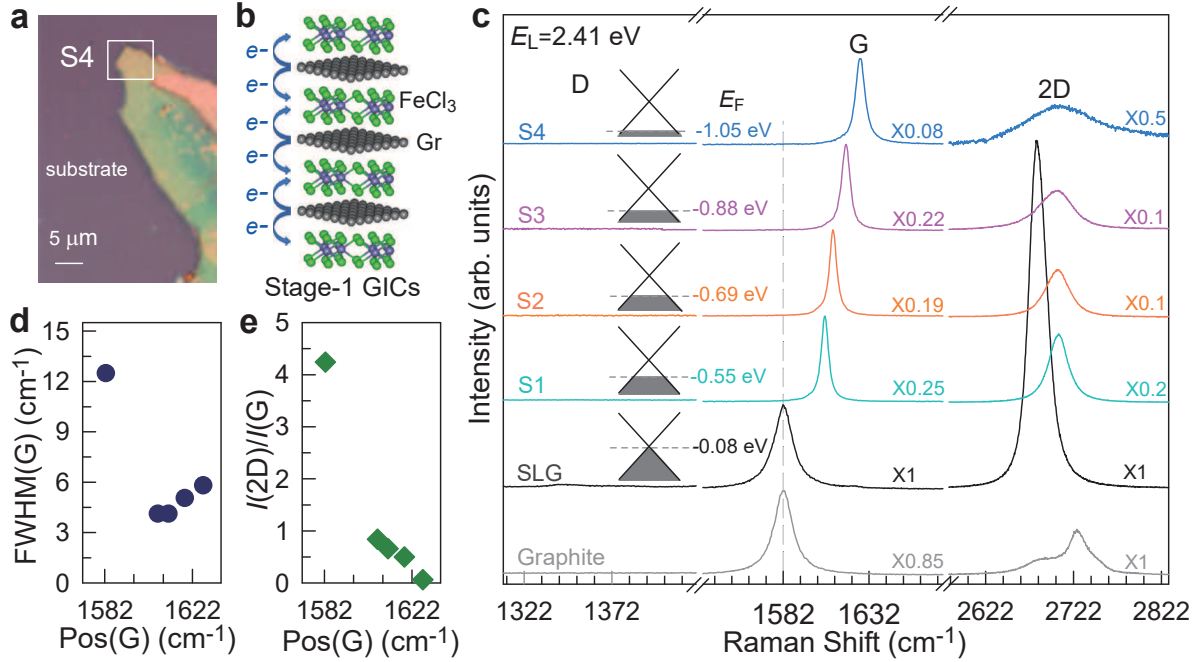}}
\caption{(a) Optical image of sample S4, with the area of interest indicated by the white box. (b) Schematic illustration of intercalation process\cite{zhao2011FeCl3}. (c) Raman spectra of samples S1-S4, with different $E_{\rm F}$, SLG ($E_{\rm F}\sim$-0.08 eV) and graphite, for $E_{\rm L}=$2.41eV. (d) FWHM(G) and (e) $I$(2D)/$I$(G) as a function of Pos(G) for SLG in samples S1-S4.}
\label{Fig.1}
\end{figure*}

We prepare 4 heavily-doped SLG samples (denoted S1-S4) by intercalating FeCl$_3$ into bulk graphite, as for Ref.\cite{zhao2011FeCl3}. FeCl$_3$ powder (Sinopharm Chemical Reagent Co., Ltd) and micromechanically exfoliated graphite flakes (Kish graphite from Graphene Supermarket) are deposited on Si covered with 90nm SiO$_2$ and positioned in the different zones of a glass tube. This is then pumped to $\sim$1.5$\times$10$^{-4}$ Torr and heated at 393K for 30mins to keep the FeCl$_3$ anhydrous. Next, the glass tube is sealed and placed in the furnace with a reaction temperature of 613K for 30h. The samples are then immediately exposed to air. Fig.\ref{Fig.1}a is a representative optical image of one sample (S4). Fig.\ref{Fig.1}b shows a schematic of FeCl$_3$-intercalated trilayer graphene, resulting in 3 individual heavily-doped SLG sandwiched by FeCl$_3$ layers.

Raman spectra are measured using a Jobin-Yvon HR800 micro-Raman system equipped with 1200 and 1800 grooves/mm gratings, coupled with a liquid-nitrogen-cooled charge coupled device (CCD) or an InGaAs array detector and a $\times$50 objective lens with a numerical aperture of 0.55. We use $E_{\rm L}$=1.16, 1.88, 2.33eV from diode-pumped solid-state lasers, 1.24$\sim$1.58 eV from a tunable continuous-wave Ti:Saphire laser, 1.96, 2.03, 2.09, 2.28eV from He-Ne lasers, 1.83, 1.92, 2.18eV from a Krypton ion laser, 2.41, 2.54, 2.62, 2.71eV from an Ar ion laser. The laser power is kept $<$2mW to avoid sample heating. During measurements at each $E_{\rm L}$, the G peak of a graphite flake with thickness $\sim$100nm is measured under the same experimental conditions to normalize the S1-S4 G peak intensity, $I$(G), for the calibrated REPs\cite{cross-section-2007-PRB,efficiency-2013-PRB}.

Fig.\ref{Fig.1}c plots the 2.41eV Raman spectra of S1-S4, not intentionally doped SLG, and graphite. In the not intentionally doped SLG, the 2D to G intensity and areas ratios are $I$(2D)/$I$(G)$\sim$4.2 and $A$(2D)/$A$(G)$\sim$8.4, respectively, indicating h doping with $E_{\rm F}\sim$-0.08eV\cite{das2008monitoring,Basko-2009-PRB}. We take this non-intentionally doped SLG as representing intrinsic SLG. For S1-S4, the Pos(G) and Pos(2D) blueshift is a signature of doping\cite{pisana2007breakdown,das2008monitoring,Ferrari-2013-NN}. $E_{\rm F}$ can be estimated by combining Pos(G), Pos(2D), FWHM(G), $I$(2D)/$I$(G) and $A$(2D)/$A$(G)\cite{pisana2007breakdown,das2008monitoring, Basko-2009-PRB,Ferrari-2013-NN} as giving $E_{\rm F}$$\sim$-0.55, -0.69, -0.88, -1.05eV, respectively.

\begin{figure*}[tb]
\centerline{\includegraphics[width=120mm]{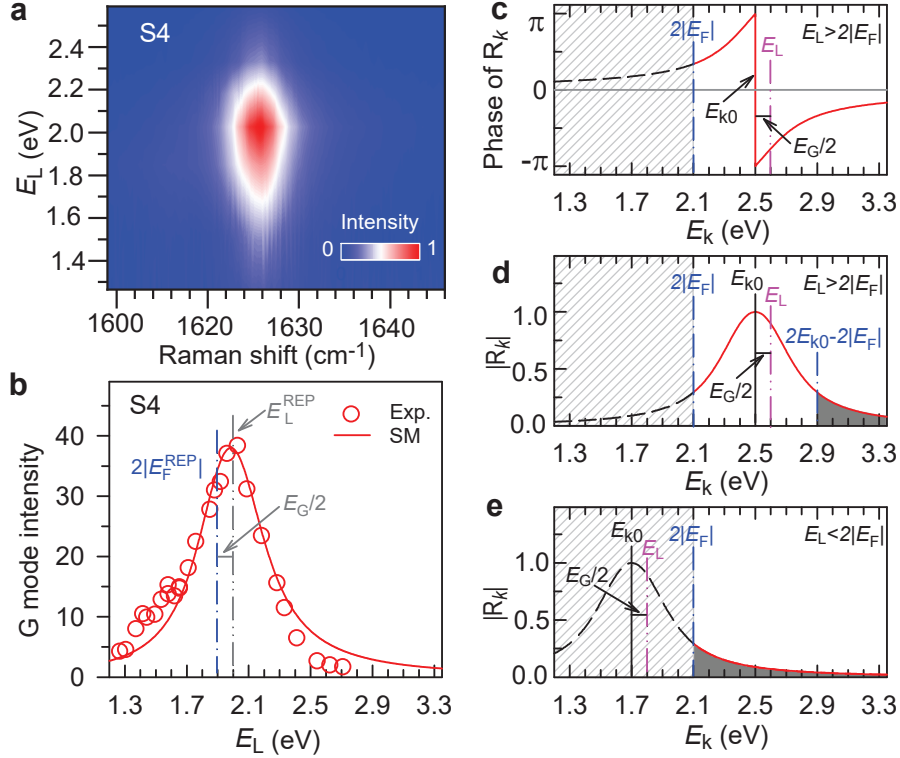}}
\caption{(a) Contour plots of $I$(G) of S4 as a function of Pos(G) and $E_{\rm L}$. (b) Experimental G REP and fit based on Eq.\ref{eq1}. Calculated (c) phase of $R_{\bm k}$ and (d) magnitude $|R_{\bm k}|$ for each pathway at $E_{\rm L}$=2.6eV and $\gamma$=0.225eV for SLG with $2|E_{\rm F}|$=2.1eV. (e) Calculated $|R_{\bm k}|$ for $E_{\rm L}$=1.8eV. The diagonal and shaded areas indicate the blocking region imposed by the Pauli exclusion principle and the pathways contributing to $I$(G).}
\label{Fig.2}
\end{figure*}

\subsection{Quantum interference and Raman excitation profile}
As $E_{\rm L}$ ranges from 1.5 to 2.7eV, the experimental $I$(G) in graphite and intrinsic SLG is almost constant when normalized to the Raman signals of calcium fluoride or cyclohexane\cite{cross-section-2007-PRB,efficiency-2013-PRB} due to the perfect cancellation of the destructive interference among different pathways\cite{Basko-NJP09-Theo,wangfeng2011Nature}. For a fixed $E_{\rm L}$, a strong increase of $I$(G) occurs as $|E_{\rm F}|$ is tuned close to $E_{\rm L}/2$\cite{wangfeng2011Nature}, due to Pauli blocking of destructive quantum interference\cite{Basko-NJP09-Theo,wangfeng2011Nature}.

Fig.\ref{Fig.2}a plots Raman measurements of S4 ($E_{\rm F}$=-1.05 eV) from 1.26 to 2.71eV over 26 steps: 1.26, 1.31, 1.37, 1.42, 1.44, 1.49, 1.53, 1.58, 1.65, 1.62, 1.65, 1.71, 1.76, 1.85, 1.88, 1.92, 1.96, 2.03, 2.09, 2.18, 2.28, 2.33, 2.41, 2.54, 2.62, 2.71eV. $I$(G) is normalized to that of bulk graphite. All the spectra in Fig.\ref{Fig.2}a have FWHM(G) as narrow as 5.8cm$^{-1}$, because e-ph scattering is forbidden by Pauli blocking once $|E_{\rm F}|$ is larger than half the G peak energy, $E_{\rm G}/2$\cite{pisana2007breakdown,das2008monitoring, zhao2011FeCl3,Ferrari-2013-NN}, therefore, we consider $I$(G) to determine the G REP in Fig.\ref{Fig.2}b. In contrast to the $E_{\rm L}$-insensitive $I$(G) in intrinsic SLG\cite{cross-section-2007-PRB,efficiency-2013-PRB}, Fig.\ref{Fig.2}b indicates that $I$(G) depends on $E_{\rm L}$, reaching a maximum for $E_{\rm L}\sim$2eV, close to $2|E_{\rm F}|$.

\begin{figure*}[tb]
\centerline{\includegraphics[width=120mm]{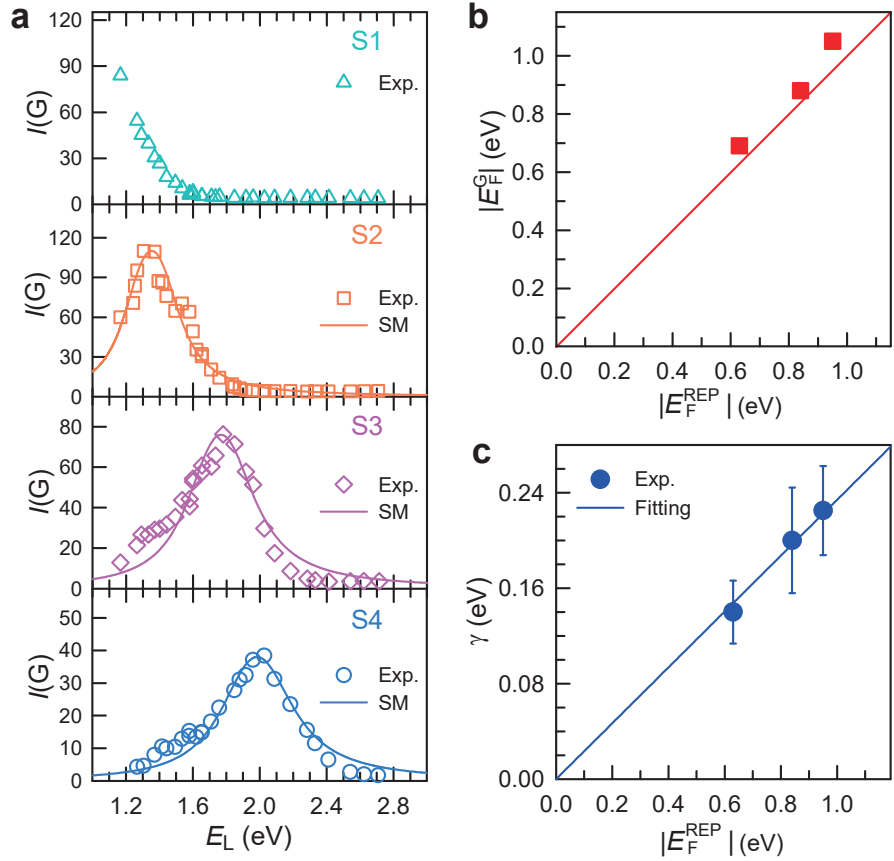}}
\caption{(a) Experimental REPs for S1-S4 along with the fitted curves based on Eq.\ref{eq1}. (b) Correlation between $|E^{\rm REP}_{\rm F}|$ and $|E^{\rm G}_{\rm F}|$. The solid line corresponds to $|E^{\rm REP}_{\rm F}|=|E^{\rm G}_{\rm F}|$. (c) $\gamma$ as a function of $|E^{\rm REP}_{\rm F}|$. The solid line is a linear fit.}
\label{Fig.3}
\end{figure*}

To explain the G REP in doped SLG, we calculate $I$(G) as a first-order Raman scattering process with e interacting only through mean-field potentials\cite{Light-Scattering-in-Solids}:
\begin{equation}
I(G)=|\sum_{\bm k}M_{\bm k}R_{\bm k}|^2
\label{eq1}
\end{equation}
\noindent where $R_{\bm k}=1/[(E_{\rm L}-E_{\bm k}+i\gamma)(E_{\rm L}-E_{\rm G}-E_{\bm k}+i\gamma)]$ is the resonance factor, $E_{\bm k}$ the vertical electronic transition energy at wavevector ${\bm k}$, $\gamma$ is the energy broadening of the excited state, comprising contributions from e-e interactions ($\gamma^{ee}$) and e-ph coupling ($\gamma^{ep}$)\cite{Basko-2009-PRB,Eb-2011-PRBPedro}, and $M_{\bm k}$ is a third-order transition matrix element (see Methods). $R_{\bm k}$ refers to one of the Raman scattering pathways\cite{wangfeng2011Nature}. When summing over ${\bm k}$, the pathways interfere with each other, which leads to a constructive or destructive effect on $I$(G), depending on the phase of allowed pathways\cite{Basko-NJP09-Theo,wangfeng2011Nature}.

We first consider a simplified model (SM) with $M_{\bm k}$ in Eq.\ref{eq1} constant for all ${\bm k}$. In this case, $I$(G)$\propto|\sum_{\bm k}R_{\bm k}|^2$. Fig.\ref{Fig.2}c,d plots the calculated phase and magnitude of $R_{\bm k}$ under $E_{\rm L}$=2.6eV and $\gamma$=0.225eV, fitted as discussed below. An abrupt phase transition (from $\pi$ to -$\pi$) is seen at $E_{\bm k_0}=E_{\rm L}-E_{\rm G}/2$, which makes the other non-resonant scattering pathways antisymmetric in phase, as referred to $E_{\bm k_0}$. For intrinsic SLG, all quantum pathways interfere destructively, leading to a weak $I$(G), as shown in Fig.\ref{Fig.1}c. However, in doped SLG, those pathways with $E_{\bm k}<2|E_{\rm F}|$ are Pauli blocked\cite{wangfeng2011Nature,zhao2011FeCl3}, as for the diagonal pattern in Figs.\ref{Fig.2}c,d with $2|E_{\rm F}|$=2.1eV. Thus, the corresponding antisymmetric pathways with $E_{\bm k}>2E_{\bm k_0}-2|E_{\rm F}|$ (shaded region in Fig.\ref{Fig.2}c) will contribute to the $I$(G) enhancement. The G REP peak occurs at $E_{\rm L}=2|E_{\rm F}|+E_{\rm G}/2$, denoted as $E^{\rm REP}_{\rm L}$. In this case, all allowed scattering pathways are in-phase. As $E_{\bm k_0}$ lies in the blocking region of $2|E_{\rm F}|$, Fig.\ref{Fig.2}e, the number of allowed in-phase pathways becomes smaller when $E_{\rm L}$ is farther from $2|E_{\rm F}|$, and the corresponding overall $I$(G) signal is weaker. Therefore, by changing $E_{\rm L}$, one can control the allowed Raman scattering pathways, creating a REP peak, Fig.\ref{Fig.2}b.

\begin{figure*}[tb]
\centerline{\includegraphics[width=100mm]{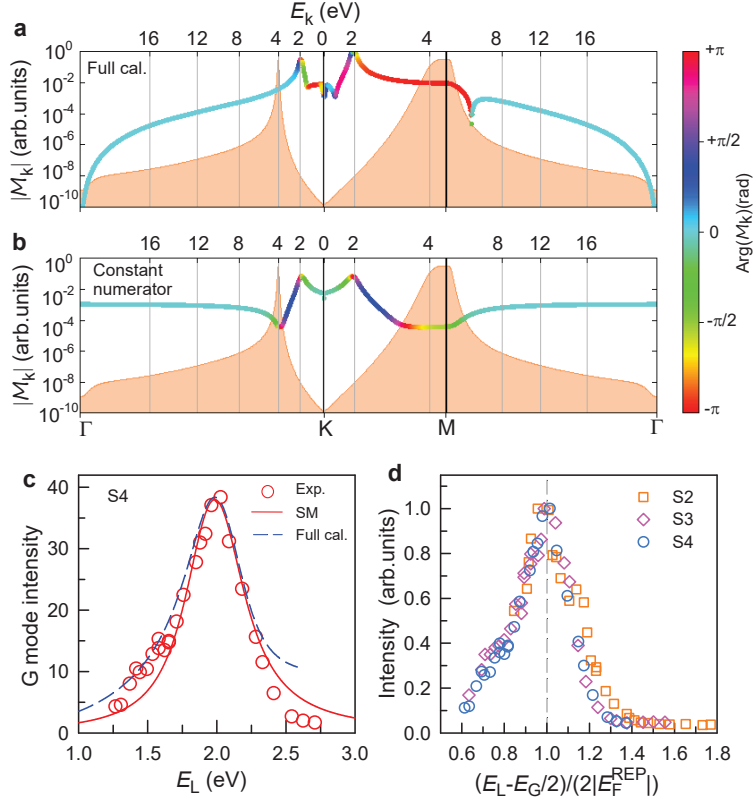}}
\caption{Absolute value (logarithmic scale) and phase (color-encoded) of $\mathcal{M}_{\bm k}$ ($\mathcal{M}_{\bm k}=M_{\bm k}R_{\bm k}$) in the high-symmetry line $\Gamma-{\rm K}-{\rm M}-\Gamma$ at $E_{\rm L}$=2eV by including $M_{\bm k}$ (a) (Full cal.), and setting the dipole and EPMEs to a constant (b) (Constant numerator), both for a constant broadening of $\gamma=0.225$eV. The shaded area represents the value of the joint density of states (JDOS) at $E_{\bm k}$. (c) Experimental REP (open circles) and theoretical REPs calculated {\it ab initio} (Full cal., dashed line) and SM (solid line). (d) Rescaled experimental REPs for S2-S4 as a function of ($E_{\rm L}-E_{\rm G}/2$)/2$|E^{\rm REP}_{\rm F}|$.}
\label{Fig.4}
\end{figure*}

Based on SM, the REP peak occurs at $E^{\rm REP}_{\rm L}=2|E_{\rm F}|+E_{\rm G}/2$. This allow us to define $E_{\rm F}$ according to the observed $E^{\rm REP}_{\rm L}$, i.e., $2E^{\rm REP}_{\rm F}=E^{\rm REP}_{\rm L}-E_{\rm G}/2$. As illustrated by the vertical lines in Fig.\ref{Fig.2}b, $|E^{\rm REP}_{\rm F}|$ is $\sim$0.95eV for S4, consistent with that extracted from the Pos(G) shift. $|E^{\rm REP}_{\rm F}|$ determines the lower bound in the ${\bm k}$ summation of Eq.\ref{eq1} to reproduce the experimental REP. The solid line in Fig.\ref{Fig.2}b is the fitted curve to the experimental REP with $\gamma$=0.225$\pm$0.04 eV, and it is larger than $\gamma\sim$0.13eV for SLG with $E_{\rm F}\sim$0.5eV on 300nm-SiO$_2$/Si\cite{Basko-2009-PRB}.

\subsection{Effects of electron-electron interaction}
Fig.\ref{Fig.3}a depicts the experimental REPs of S1-S4 measured by varying $E_{\rm L}$ from 1.26 to 2.71eV. Due to the increasing E$_F$ from S1,-0.55eV, to S4,-1.05eV, the corresponding $E^{\rm REP}_{\rm L}$ blue-shifts. Based on the experimental $E^{\rm REP}_{\rm L}$, $E^{\rm REP}_{\rm F}$ of S2-S4 can be determined as $\sim$-0.63, -0.84, -0.95eV, respectively, consistent with $|E_{\rm F}|$ from Pos(G), Pos(2D), $I$(2D)/$I$(G), $A$(2D)/$A$(G), Fig.\ref{Fig.3}b.

The fitted $\gamma$ from REPs of S2-S4 with Eq.\ref{eq1} increase monotonically with $|E^{\rm REP}_{\rm F}|$, i.e., $\gamma=0.234|E^{\rm REP}_{\rm F}|$, Fig.\ref{Fig.3}c. $\gamma$ is related to the broadening of excited states, due to the interactions with elementary excitations, such as doping-induced e/h, ph, and defects\cite{Basko-2009-PRB}. Since S2-S4 are defect-free, as shown from the absence of the D peaks in Fig.\ref{Fig.1}c, we can write $\gamma=\gamma^{ee}+\gamma^{ep}$\cite{Basko-2009-PRB}. $\gamma^{ee}$ denotes the e-e scattering rate, which increases as more e/h are added to SLG. $\gamma^{ee}=2|E_{\rm F}|f(e^2/2\varepsilon_0\varepsilon h\nu_F)$, with $\varepsilon_0$, $\varepsilon$, $h$, $\nu_F$ vacuum permittivity, dielectric constant, Planck constant, and Fermi velocity, respectively\cite{Basko-2009-PRB}. From Ref.\cite{Basko-2009-PRB}, we get $f\sim$0.09\cite{zhao2011FeCl3}, i.e., $\gamma^{ee}=0.18|E_{\rm F}|$, smaller than the fitted slope$\sim$0.234 in Fig.\ref{Fig.3}c. On the other hand, $\gamma^{ep}$ does not depend explicitly on $E_{\rm F}$. From Ref.\cite{Eb-2011-PRBPedro}, $\gamma^{ep}$ is dispersive with $E_{\bm k}$ as $\gamma^{ep}=0.021E_{\bm k}-0.0034$. This must be considered when summing ${\bm k}$ in Eq.\ref{eq1}. However, since the slope of the dispersion, $\sim$0.021, is $<<$1, $\gamma^{ep}$ can be approximated to $\gamma^{ep}\sim0.042|E_{\rm F}|$ (see Methods). So, the overall slope of $\gamma$ is $\sim$0.22, in agreement with the fit in Fig.\ref{Fig.3}c. Thus, the REP energy broadening in doped SLG comes mainly from enhanced e-e interactions.

We now use {\it ab initio} density functional and many-body perturbation theory to calculate the full Raman scattering matrix-element $\mathcal{M}_{\bm k}$ ($\mathcal{M}_{\bm k}=M_{\bm k}R_{\bm k}$) on the independent-particle level (see Methods). Fig.\ref{Fig.4}a plots the dispersion of the full scattering matrix element and compares it with SM (Fig.\ref{Fig.4}b) for $E_{\rm L}$=2eV. The constant dipole and e-ph matrix elements (EPMEs) are individually set to the square root of their average modulus, which is squared taken over bands, polarizations and the resonant ${\bm k}$-point surface. The full matrix element is much more dispersive than in SM, which only shows the two peaks at the surface of resonant ${\bm k}$-points. We attribute the more pronounced resonance peaks in the full calculation to two effects: (i) the underestimation of the decay of the dipole and EPMEs far away from the BZ edge at the K-point in SM, and (ii) the finite scattering matrix element at the K-point in SM vanishes in the full calculations, because the phase of the numerator in Eq.\ref{eq1} is constant, rather than rotating around K, in line with approximate angular momentum conservation\cite{Basko-NJP09-Theo,Sven-2017-PRB}. Thus, SM captures the largest part of the physics through its almost double-resonant structure, but leads to a quantitative underestimation of resonance effects. The full calculated REP for S4 is normalized to the experimental data, Fig.\ref{Fig.4}c. This is slightly broadened as compared to SM, improving the agreement on the red-side of the resonance peak in the experimental REP.

The fitted $\gamma$ from REPs is linear with $|E^{\rm REP}_{\rm F}|$, as indicated in Fig.\ref{Fig.3}c. The quantum interference amongst Raman scattering pathways mainly depends on $E_{\bm k}$ away from $2E_{\rm L}-2|E_{\rm F}|-E_{\rm G}$ ($E_{\rm L}>2|E_{\rm F}|$) or $2|E_{\rm F}|$ ($E_{\rm L}<2|E_{\rm F}|$) (Fig.\ref{Fig.2}c, d), which can be finely tuned by $E_{\rm L}$ or $E_{\rm F}$. The established relations of $2|E^{\rm REP}_{\rm F}|=E^{\rm REP}_{\rm L}-E_{\rm G}/2$ can now be used to rescale $E_{\rm L}$ of the REPs in Fig.\ref{Fig.3}a. The corresponding rescaled REPs, i.e., $I$(G) as a function of $(E_{\rm L}-E_{\rm G}/2)/2|E^{\rm REP}_{\rm F}|$, are in Fig.\ref{Fig.4}d for S2-S4. The 3 rescaled REPs show a similar profile, although their $E^{\rm REP}_{\rm F}$ is different, confirming the linear dependence of $\gamma$ on $E_{\rm F}$, as for the Raman measurements in Fig.\ref{Fig.3}a.

In summary, we carried out a systematic experimental study of the quantum interference effects on the Raman scattering pathways of the G mode in doped SLG. By adjusting $E_{\rm L}$ over 26 individual energies between 1.2 and 2.7eV, we controlled the number of Raman scattering pathways, in order to enhance or attenuate $I$(G), reaching a maximum for $E_{\rm L}=2|E_{\rm F}|+E_{\rm G}/2$. The dispersive $\gamma$ can be fitted from the experimental REPs, and is linearly related to $E_{\rm F}$, with the main contribution dominated by e-e interactions. REPs can be rescaled by $E^{\rm REP}_{\rm L}=2|E^{\rm REP}_{\rm F}|+E_{\rm G}/2$. Thus REP is a powerful tool for probing electronic interactions.

\section{Methods}
{\it ab initio} calculations of the Raman matrix elements are done as for Refs.\cite{Sven-2017-PRB,Sven-2020-SciAdv}. The SLG band structure, e-light, and the screened e-ph matrix elements are obtained from density functional (perturbation) theory, with the PWscf code from Quantum ESPRESSO\cite{Giannozzi-2009-QE,Giannozzi-2017-QE} using a plane-wave basis set with an energy cutoff of 80Ry. An ultrasoft pseudopotential is used to describe the e-ion interaction, while the mean-field exchange-correlation potential is approximated on the level of the generalized gradient approximation in the parametrization by Perdew, Burke, and Ernzerh of \cite{Perdew-PBE}. A vacuum spacing of 14\AA~separates periodic SLG copies, with the relaxed value of 2.46\AA~for the lattice constant\cite{Sven-2017-PRB}. A uniform 60$\times$60$\times$1 ${\bm k}$-point mesh is used to sample the first BZ in a self-consistent calculation for the ground state density and potential, and for the calculation of the change of the self-consistently screened lattice potential with the ph displacement. Due to the SLG semi-metallic nature, the latter requires a thermal smearing, for which the electronic states are populated according to a Fermi-Dirac distribution with temperature corresponding to 0.002Ry. To obtain converged results for the Raman intensity, the electronic $\pi$ and $\pi^*$ and the optical and e-ph matrix elements are interpolated to a dense 480$\times$480$\times$1 ${\bm k}$-point mesh using maximally localized Wannier functions from a coarse 12$\times$12$\times$1 ${\bm k}$-point grid, as implemented in the Wannier90~\cite{Mostofi-2014-W90} and EPW codes\cite{Giustino-2007-EPW,Ponce-2016-EPW}. The full {\it ab initio} calculation goes beyond the approximation of retaining the almost double-resonant term in Eq.\ref{eq1}, for which $M_{\bm k}=d^i_{\bm k,\pi^*\pi} (g^{\lambda}_{\bm k,\pi^*\pi^*}-g^{\lambda}_{\bm k,\pi\pi}) (d^j_{\bm k,\pi^*,\pi})^*$, with $d^i_{\bm k,\pi^*\pi}$ denoting the $i$-th component of the dipole matrix element and $g^{\lambda}_{\bm k,nn}$ the diagonal screened e-ph matrix element for band $n=\pi,\pi^*$ for ph polarization $\lambda=x,y$. Instead, it includes all possible time orderings of the independent-particle three-particle correlation\cite{Sven-2020-SciAdv}.

For simplicity, $E_{\rm G}$ is ignored in Eq.\ref{eq1}. Then, $\Delta=E_{\bm k}-E_{\bm L}=E_{\bm k}-E_{\bm k_0}$, where $E_{\bm L}=E_{\bm k_0}$ applies since $E_{\rm G}$ is not included. Thus, the dispersive $\gamma=\beta E_{\bm k}$ can be arranged as $\gamma=\beta \Delta +\gamma_0$, with $\gamma_0=\beta E_{\bm k_0}$. Thus, the sum over ${\bm k}$ is equivalent to integrating over $\Delta$, giving:
\begin{equation}
\begin{split}
\begin{aligned}
I(G)&=\left|\sum_{\bm k}R_{\bm k}\right|^2\\
&=\left|\int \frac{1}{[\Delta +i(\beta \Delta +\gamma_0)]^2} d \Delta\right|^2\\
&=\frac{1}{(\beta^2 +1)^2}\frac{1}{(\Delta +\frac{\beta \gamma_0}{\beta^2 +1})^2+(\frac{\gamma_0}{\beta^2 +1})^2}
\end{aligned}
\end{split}
\label{eq2}
\end{equation}
The maximum $I(G)$ [$I(G)^{max}$] is $1/{\gamma_0^2}$ and the FWHM of the profile is $2\gamma_0/(\beta^2 +1)$. Since $\beta=$ 0.021$<<$1\cite{Eb-2011-PRBPedro}, the FWHM is approximated by $2\gamma_0=2\beta E_{\bm k_0}$, only related to $\gamma$ at $E_{\bm k_0}$. Eq.\ref{eq2} then simplifies to $I(G)=1/({\Delta^2 +\gamma_0^2})$, which is exactly the same as setting $\gamma=$ as constant:
\begin{equation}
I(G)=\left|\int \frac{1}{(\Delta +iC)^2} d \Delta\right|^2=\frac{1}{\Delta^2 +C^2}
\label{eq3}
\end{equation}
with C replaced by $\gamma_0$. This suggests that the constant $\gamma$ widely used in the literature\cite{wangfeng2011Nature,Liu-2013-nanolett} comes from $\gamma_0$ defined at $E_{\bm k_0}$. The area of the profile can be further obtained by multiplying $I(G)^{max}$ by the FWHM, which is $2/{\gamma_0}$.

In doped SLG, $E_{\bm k_0}=2|E_{\bm F}|$. Thus, $\gamma^{ep}=0.021E_{\bm k}-0.0034$ gives $\gamma^{ep}=0.042|E_{\rm F}|-0.0034\sim0.042|E_{\rm F}|$.

\noindent {\bf Competing interests:} The authors declare no competing financial interest.

\begin{acknowledgement}
We thank W. L. Ma for fruitful discussions. We acknowledge support from the National Natural Science Foundation of China (Grant Nos. 12127807, 12004377, and 12174381), CAS Key Research Program of Frontier Sciences (Grant Nos. ZDBS-LY-SLH004 and XDPB22), CAS Project for Young Scientists in Basic Research (YSBR-026), and the National Reasearch Fund (FNR) Luxembourg, project “RESRAMAN” (Grant No. C20/MS/14802965), EU Graphene Flagship, ERC Grants Hetero2D, GIPT, EU Grants GRAP-X, CHARM, EPSRC Grants EP/K01711X/1, EP/K017144/1, EP/N010345/1, EP/L016087/1, EP/V000055/1, EP/X015742/1.
\end{acknowledgement}





\providecommand{\latin}[1]{#1}
\makeatletter
\providecommand{\doi}
  {\begingroup\let\do\@makeother\dospecials
  \catcode`\{=1 \catcode`\}=2 \doi@aux}
\providecommand{\doi@aux}[1]{\endgroup\texttt{#1}}
\makeatother
\providecommand*\mcitethebibliography{\thebibliography}
\csname @ifundefined\endcsname{endmcitethebibliography}
  {\let\endmcitethebibliography\endthebibliography}{}

\end{document}